\documentclass[prb,twocolumn]{revtex4}
\usepackage{amsmath}
\usepackage{amsfonts}
\usepackage{amssymb}
\usepackage{graphicx}

\begin{document}

\title{Single Qubit Rabi Oscillation Decohered by Many Two-Level Systems}
\author{Ren-Shou Huang}
\affiliation{Ames Laboratory, Iowa State University, Ames, IA 50011, USA}
\author{Viatcheslav Dobrovitski}
\affiliation{Ames Laboratory, Iowa State University, Ames, IA 50011, USA}
\author{Bruce Harmon}
\affiliation{Ames Laboratory, Iowa State University, Ames, IA 50011, USA}
\date{\today}
\bibliographystyle{apsrev}

\begin{abstract}
Recent experiments on Josephson junction qubits have suggested the existence in the tunnel barrier of bistable two level fluctuators that are responsible for decoherence and 1/f critical current noise. In this article we treat these two-level systems as fictitious spins and investigate their influence quantum mechanically with both analytical and numerical means. We find that the Rabi oscillations of the qubit exhibit multiple stages of decay. New approaches are established to characterize different decoherence times and to allow for easier feature extraction from experimental data. The Rabi oscillation of a qubit coupled to a spurious resonator is also studied, where we proposed an idea to explain the serious deterioration of the Rabi osillation amplitude.
\end{abstract}

\maketitle

\section{Introduction}
Quantum computing with superconducting circuits has become one of the most anticipated approaches for the actual realization of quantum computation. However, like other solid state qubit candidates, while enjoying the advantage of scalability and the ease of manufacturing, it suffers from short coherence time. This is mainly because the qubit states used in these devices are macroscopic quantum states, in contrast to the microscopic states used in NMR, ion traps, cavity QED, etc., which obviously have fewer intrinsic degrees of freedom that are open to perturbation from the external environment.

Based on recent experiments, it has been shown that the primary noise that dephases a superconducting qubit has an 1/f like spectrum in a given low frequency range.\cite{Astafiev04} Although the mechanism of this process is still unknown, it is speculated to be related to the critical current fluctuations in the Josephson junction due to its couplings to many two-level systems\cite{Simmonds04}. There are several possible explanations. The one we will focus on is that the trapped charge on the impurities in the tunnelling barrier of the junction block the junction area by Coulomb repulsion, thus modifying the critical current and the Josephson energy\cite{vanHarlingen04,Rogers84,Rogers85}.

There have been several theoretical studies with similar models used to investigate the decoherence by evaluating the perturbation on the Rabi oscillations of the qubit\cite{Paladino02,Ku04}. In this article we analyze analytically and numerically the details of the Rabi oscillations for a qubit weakly coupled to an environment of two level systems. The time dependence of decoherence reveals different characteristics in the short and long time regimes, which can be used to analyze experimental data and extract information about the actual qubit energy fluctuation. Instead of trying to solve the problem exactly at once, in Sec. II, we started from an approximated model of a qubit dephased by many spins, which contains all the essential features where we can understand them in Sec. III, IV, V qualitatively and quantitatively by both analytical and numerical means. Later in Sec. VI we introduce other elements such as exchange interactions between the qubit and the spins whose energy is close to the qubit's. This results in the relaxation effect and sometimes the complete destruction of coherence when they are exactly on resonance, as explained in Sec. VII.

\section{Theoretical Model}
The theoretical model of the system we consider is basically a qubit coupling to many other spins while undergoing microwave driven Rabi oscillation. The Hamiltonian is
\begin{eqnarray}
\mathcal{H} & = & \mathcal{H}_\mathrm{qb}+\mathcal{H}_\mathrm{spins}+\mathcal{H}_\mathrm{int}+\mathcal{H}_\mathrm{Rabi},\nonumber\\
\mathcal{H}_\mathrm{qb} & = & \frac{\Omega}{2}\sigma^z,\nonumber\\
\mathcal{H}_\mathrm{spins} & = & \sum_k\frac{\omega_k}{2}\tau_k^z,\nonumber\\
\mathcal{H}_\mathrm{int} & = & \sum_kA_k\tau_k^z\sigma^z,\nonumber\\
\mathcal{H}_\mathrm{Rabi} & = & \alpha(\sigma^+e^{-i\Omega't}+\sigma^-e^{i\Omega't}).\label{eq:H}
\end{eqnarray}
The $\sigma$ and $\tau$ in the Hamiltonian are Pauli matrices. $\Omega$ and $\omega_k$ are the energy splittings of the qubit and the spins. $A_k$ are the qubit-spin coupling constants. $2\alpha$ is the Rabi frequency and $\Omega'$ is the microwave frequency.

In the charge trapping picture, the $\omega_k$ are given by $2\Delta-\epsilon_k$, where $\Delta$ is the gap energy of the superconductor and $\epsilon_k$ is the trapping energy of the impurity in a Hubbard model\cite{Kozub03}. When $\epsilon_k>0$, the trap is attractive. The $\omega_k$ here represent the energy differences of the processes where a Cooper pair breaks up and one of its electrons hops onto the impurity. Such attractive impurities have been found, for instance, in the so called $D^X$-centers in semiconductors formed by substitutional dopants in GaAs and AlGaAs alloys\cite{Chadi89,Shi96}. We assume the density of the traps is sufficiently low so that they don't interact with each other. Also the junction barrier is thought to be so thin that electrons only hop onto one single trap instead of a chain of traps.

$A_k$ means the amount of qubit energy being changed when the $k$th trap is occupied or deoccupied. The $\mathcal{H}_\mathrm{Rabi}$ represents the Hamiltonian of the qubit under the driving microwave at frequency $\Omega'$ with the rotating wave approximation.

For the interactions between the qubit and the spins we only include the terms where the spins perturb the energy splitting of the qubit. Here we omit the qubit flipping in the first part of this article because we want to investigate solely the dephasing caused by the many-spin system. Since the charge trapping processes only modulate the Josephson energy, this can be viewed, for example, as a model for a Cooper pair box at the charge degeneracy point. However, this model is not limited to only this picture. It is generally applicable to all qubit systems with dephasing sources approximated as quantum two-level systems.

\section{Analytical Solution}
The Hamiltonian in Eq.~(\ref{eq:H}) is exactly solvable when the number of spins is infinity\cite{Dobrovitski03-1}. First we apply an unitary transformation,
\begin{equation}
U(t)=\mathrm{exp}\left(i\frac{\Omega't}{2}\sigma^z+i\sum_k\frac{\omega_kt}{2}\tau_k^z\right).
\end{equation}
So that in this rotating frame the effective Hamiltonian is
\begin{equation}
\tilde{\mathcal{H}}=\frac{\Omega-\Omega'}{2}\sigma^z+\sum_kA_k\tau_k^z\sigma^z+\alpha\sigma^x.\label{eq:H_eff}
\end{equation}
Define $B=\sum_kA_k\tau_k^z$. Since $\tau_k^z$ commute with the Hamiltonian, they are all constants of motion and their values have their own thermal distributions.  The qubit energy splitting is modified by its coupling to the spins and becomes $\Omega+2\langle B\rangle$. Since this new energy splitting is the one that is measured in experiments, to drive Rabi oscillation, we let $\Omega'=\Omega+2\langle B\rangle$. Thus Eq.~(\ref{eq:H_eff}) is simplified to
\begin{equation}
\tilde{\mathcal{H}}=(B-\langle B\rangle)\sigma^z+h\sigma^x\equiv B'\sigma^z+\alpha\sigma^x.
\end{equation}
The evolution of the physical quantity $\sigma^z$ can be calculated,
\begin{equation}
\langle\sigma^z(t)\rangle=\mathrm{Tr}[ e^{i\tilde{\mathcal{H}}t}\sigma^ze^{-i\tilde{\mathcal{H}}t}\rho_0].\label{eq:sigma_t}
\end{equation}
The initial density matrix is
\begin{equation}
\rho_0=|\downarrow\rangle\langle\downarrow|\otimes e^{-\beta\sum_k\frac{\omega_k}{2}\tau_k^z},\label{eq:rho0}
\end{equation}
so that the qubit is in the $|\downarrow\rangle$ state and the spins are in thermal equilibrium at temperature $T=1/k_B\beta$. This is close to the real state of the system when the temperature $T\ll\Omega$. Note that the time evolution operator can be rewritten as
\begin{equation}
e^{i\tilde{\mathcal{H}}t}=\cos\sqrt{\alpha^2+B'^2}t+i\frac{\alpha\sigma^x+B'\sigma^z}{\sqrt{\alpha^2+B'^2}}\sin\sqrt{\alpha^2+B'^2}t.
\end{equation}
Because the initial state is separable, the calculation of Eq.~(\ref{eq:sigma_t}) ends up with calculating terms like $\langle B'^n\rangle=\mathrm{Tr}[B'^n\rho_0]$. Since in this approximated model $\tau_k^z$ are all independent random variables governed only by temperature, for a very large number of spins, the value of $B'$ is given by a Gaussian distribution with
\begin{equation}
\langle B'\rangle=0,\quad\langle B'^2\rangle=\sum_k\frac{A_k^2}{\cosh^2\frac{\beta\omega_k}{2}}\equiv \delta\Omega^2.\label{eq:b_beta}
\end{equation}
The quantity $\delta\Omega>0$ is a measure of fluctuation of the qubit energy splitting due to finite temperature. We can then obtain the exact solution in a closed form
\begin{equation}
\langle\sigma^z(t)\rangle=-1+\frac{2\alpha^2}{\sqrt{2\pi}\delta\Omega}\int_{-\infty}^\infty dB'\frac{\sin^2\sqrt{\alpha^2+B'^2}t}{\alpha^2+B'^2}e^{-\frac{B'^2}{2\delta\Omega^2}}.\label{eq:sz}
\end{equation}
Although the spins considered here are coupled to a thermal reservoir which gives rise to the Gaussian distribution of the variable $B'$, the same formalism still applies when the Gaussian distribution in Eq.~(\ref{eq:sz}) is replaced by any arbitrary distribution for other mechanisms that govern the fluctuations of the two-level systems, and the precise definition of $\delta\Omega$ will have to change with it.
\begin{figure}
\centering
\includegraphics[scale=0.6]{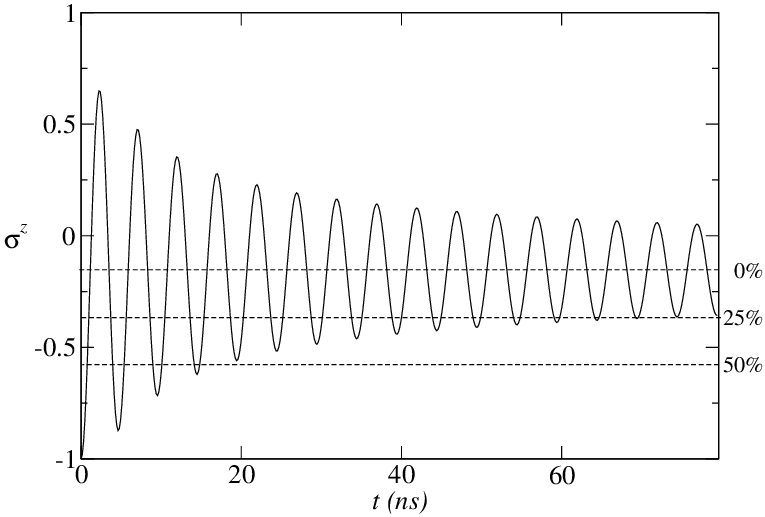}
\caption{Time evolution of the qubit decohered by many spins undergoing Rabi oscillation by a coherent microwave source, $\langle\sigma^z(t)\rangle$. The Rabi frequency $2\alpha/2\pi=200$MHz, and $\delta\Omega$ is generated from a special case where all spin energy splittings are chosen for simplification to be $\omega_k/2\pi=1$GHz and $\sqrt{\sum_kA_k^2}/2\pi=50$MHz at the temperature $T=150$mK. For $\Omega/2\pi=10$GHz, these parameters correspond to $\delta\Omega/\Omega\lesssim 0.005$. Notice that the amplitude of the oscillation envelope is already down to $50\%$ at $t\sim 20$ns, but it reaches $25\%$ only after $t=80$ns, while an exponential decay should reach $25\%$ around $t\sim 40$ns.}\label{fig:sz}
\end{figure}

\section{Decoherence}
Since this integral in Eq.~(\ref{eq:sz}) is difficult to carry out, first we look at a special case. When $T\rightarrow 0$, $\delta\Omega\rightarrow 0$. The Gaussian distribution function goes to a delta function, which means no thermal fluctuation. In this limit
\begin{equation}
\lim_{T\rightarrow 0}\langle\sigma^z(t)\rangle=-\cos2\alpha t,
\end{equation}
which is the ideal Rabi oscillation, because every spin is frozen to its own ground state. But as soon as we turn up the temperature, when the value $B$ is allowed to fluctuate, the oscillation now has a decay pattern(see in Fig.~(\ref{fig:sz})). Notice that the oscillation lasts much longer than an ordinary exponential decay but has a rather drastic decrease of amplitude in the beginning.

The Fourier transform of Eq.~(\ref{eq:sz}) is
\begin{eqnarray}
\tilde{\sigma}^z(\omega) & = & \int dte^{i\omega t}\langle\sigma^z(t)\rangle\nonumber\\
& \sim & \int dt\int_{-\infty}^\infty dB'\frac{e^{i(\omega\pm 2\sqrt{\alpha^2+B'^2})t}e^{-\frac{B'^2}{\delta\Omega^2}}}{\alpha^2+B'^2}\nonumber\\
& \sim & \frac{1}{\omega}\frac{1}{\sqrt{\omega^2-4\alpha^2}}\mathrm{exp}\left(-\frac{\omega^2-4\alpha^2}{8\delta\Omega^2}\right),\label{eq:FTsz}
\end{eqnarray}
for $|\omega|\geq2\alpha$. The meaning of Eq.~(\ref{eq:FTsz}) is clearer if we define $\omega'^2=\omega^2-4\alpha^2$ and rewrite it as
\begin{equation}
\tilde{\sigma}^z(\omega')\sim\frac{1}{\sqrt{\omega'^2+4\alpha^2}}\frac{1}{\omega'}e^{-\frac{\omega'^2}{8\delta\Omega^2}}.
\end{equation}
\begin{figure}
\centering
\includegraphics[scale=0.6]{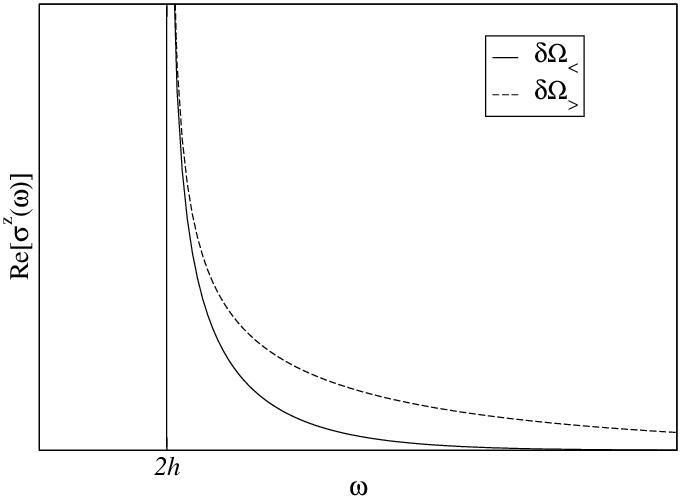}
\caption{Fourier spectrum $\tilde{\sigma}^z(\omega)$ near the Rabi frequency $\omega=2\alpha$. We can use the width of the peak to estimate the decoherence time of the Rabi oscillation. $\delta\Omega_>$ and $\delta\Omega_<$ are the larger and the smaller values of two random $\delta\Omega$. For larger $\delta\Omega$ we have a wider width and thus shorter decoherence time and vice versa. For frequency $\omega\gg2\alpha+2\delta\Omega$ the tail of the peak is dominated by the Gaussian term, therefore we can expect a Gaussian decay of $\langle\sigma^z(t)\rangle$ when $t\ll \hbar/\delta\Omega$. While the frequency $\omega$ is very near $2\alpha$, the peak goes like $(\omega-2\alpha)^{-1/2}$, which implies a very slow decay at the long time limit.}\label{fig:sz_w}
\end{figure}

Now the width of the peak $\omega'=0$, \emph{i.e.} $\omega=2\alpha$, is controlled by the Gaussian function with width $2\delta\Omega$, as shown in Fig.~(\ref{fig:sz_w}). When $\delta\Omega\rightarrow 0$, the spectrum becomes a delta function at the Rabi frequency $\omega=2\alpha$. At high frequency the Fourier spectrum is dominated by the Gaussian term, which means that $\langle\sigma^z(t)\rangle$ has a Gaussian decay in the short time limit. The singularity at the Rabi frequency $\omega=2\alpha$ implies a very slow decay in the long time limit.  The result agrees with the that obtained for Eq.~(\ref{eq:sz}) in the limit when $\alpha\gg\delta\Omega$, the oscillation envelope of $\langle\sigma^z(t)\rangle$ is given by $\sigma^z(0)[1+(2t\delta\Omega^2/\alpha)^2]^{-\frac{1}{4}}$, where the evolution begins with a Gaussian (quadratic) damping then changes to a slow power law decay of $\sim\sqrt{\alpha/\delta\Omega^2t}$\cite{Dobrovitski03-1}.

Since the mapping from $\omega$ to $\omega'$ is not linear, the Gaussian width $2\delta\Omega$ in $\omega'$ is approximately translated into a width $2(\sqrt{\alpha^2+\delta\Omega^2}-\alpha)$ in $\omega$. Thus we can define the decoherence time as
\begin{equation}
\frac{1}{T_\phi}\equiv2(\sqrt{\alpha^2+\delta\Omega^2}-\alpha).\label{eq:t_phi}
\end{equation}
It shouldn't be a surprise that the decoherence time of a Rabi oscillation depends on the Rabi frequency. In the limit of small Rabi frequency $\alpha\ll\delta\Omega$, $1/T_\phi\rightarrow 2\delta\Omega$, which means the slower the Rabi oscillation, the more it is affected by dephasing. Also the physical meaning of $T_\phi$ is similar to its counter part in an exponential decay, $T_2^*$, since both of them are the product of pure dephasing and do not involve relaxation.

\begin{figure}[!tp]
\centering
\includegraphics[scale=0.6]{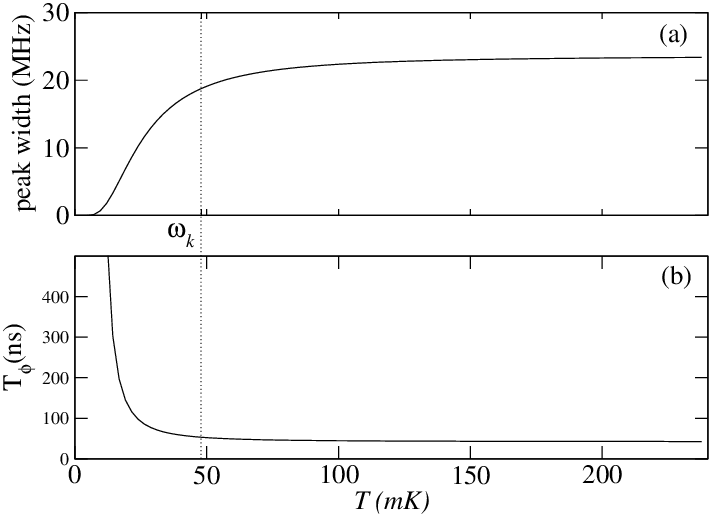}
\caption{Temperature dependence of the peak width and the decoherence time. (a) is the direct plot of Eq.~(\ref{eq:t_phi}) using the same parameters as those in Fig.~(\ref{fig:sz}). (b) is the decoherence time, which is the inverse of (a). Both the width and decoherence time saturate when $k_BT>\hbar\omega_k$.}\label{fig:dephase}
\end{figure}

The temperature dependence of the peak width and the decoherence time is plotted in Fig.~(\ref{fig:dephase}). Notice the saturation of decoherence time is due to the saturation of $\delta\Omega$, thus the spins with $|\omega_k|=|2\Delta-\epsilon_k|<k_BT$ is experimentally unfavorable. If the electron trapping picture is correct, then better tunnel junction barrier materials or fabrication procedures should be sought out that doesn't contain electron traps that have attractive trap energy close to the superconductor gap energy.

\section{Numerical Simulation}
To investigate more realistic situations we are required to solve the time-dependent Schr\"odinger equation numerically. Since the dimension of the Hilbert space grows exponentially with the number of spins, it is necessary to use an efficient and accurate method to propagate the dynamics of the system for long times. We adopt the Chebyshev expansion method\cite{Dobrovitski03-2} to carry out this task. For initialization, a randomly generated state is propagated with the Hamiltonian $\mathcal{H}_\mathrm{qb}+\mathcal{H}_\mathrm{spin}$ in an amount of imaginary time, $i\beta$, instead of real time, and then normalized, so that the initial states have a Gibbs distribution. Later the normalized initial state is propagated in real time with Hamiltonian $\mathcal{H}$, or equivalently $\tilde{\mathcal{H}}$.

First let us start with the simplified model previously considered analytically. The coupling parameters $|A_k|$ are randomly chosen between $10\sim30$MHz, and the $|\omega_k|$ of total 14 spins are uniformly distributed per decade from 2MHz to 10GHz. Since the chosen qubit bare energy splitting is $\Omega=2\pi\times10$GHz, much higher than the temperature, the initial state here is basically the same as Eq.~(\ref{eq:rho0}).

\begin{figure}
\centering
\includegraphics[scale=0.6]{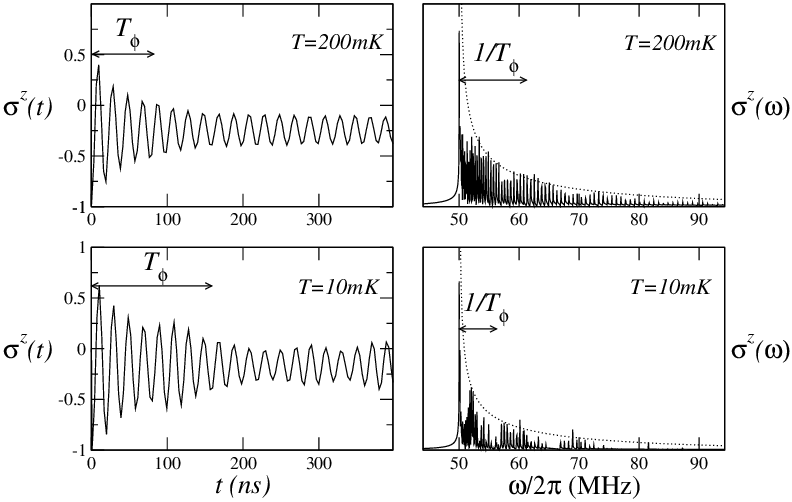}
\caption{Numerical simulation for the Rabi oscillation of one qubit coupled with 14 spins at $T=200$mK and $T=10$mK. The left column are the real time Rabi oscillations, and the right column are their Fourier transforms. The dotted lines in the right column are the exact Fourier transform obtained in Eq.~(\ref{eq:FTsz}) for the limit of infinite number of spins. $T_\phi$, calculated using Eq.~(\ref{eq:t_phi}), are 84ns and 165ns respectively. Notice that the small bumps in the oscillation envelope in the $T=10$mK are due to the effects of finite spins and some spins are frozen.}\label{fig:tdep}
\end{figure}
Fig.~(\ref{fig:tdep}) shows the results at $T=200$mK and $T=10$mK. The graphs in general look like Fig.~(\ref{fig:sz}) and Fig.~(\ref{fig:sz_w}). For the same set of parameters, higher temperature obviously gives faster and larger reduction of the oscillation amplitude. From the graph we can see that the right wing of the peak in the Fourier spectrum actually consists of many small peaks due to the finite amount of spins. The physical meaning of $T_\phi$, illustrated in the real time evolution graphs, is to characterize the process of the initial reduction of the amplitude, in this non-interacting spin case, a Gaussian decay. The much slower power-law decay that follows is just a general symptom of the coupling to many-spin systems and contains the information about dephasing only in its amplitude which was left over by the initial reduction. The rate of power-law decay cannot be used to extract any useful information. This special feature may well explain part of the reason for the problem of seriously reduced Rabi oscillation amplitude which plagues many experiments.

\section{Dephasing plus Relaxation}
The relaxation of the qubit by a many-spin system can be described by adding a term to $\mathcal{H}_\mathrm{int}$, so that now
\begin{equation}
\mathcal{H}_\mathrm{int}=\sum_kA_k\tau_k^z\sigma^z+\sum_k\lambda_k(\tau_k^+\sigma^-+\tau_k^-\sigma^+).
\end{equation}
This kind of processes has been carefully studied by Paladino \emph{et. al.}\cite{Paladino02}. The relaxation effect caused by spins with energy splittings close to the qubit's alone in general gives the Rabi oscillation an exponential decay through the relaxation induced dephasing.
\begin{figure}
\centering
\includegraphics[scale=0.6]{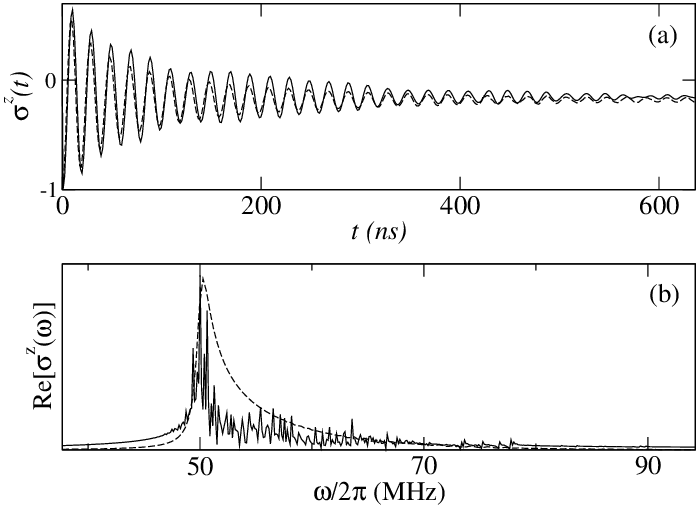}
\caption{Rabi oscillations of a qubit dephased and relaxed by a many-spin system. The parameters are the same as those used in the $T=10$mK graph in Fig.~(\ref{fig:tdep}). The only additional parameter, the relaxation time $T_1$, calculated from Fermi's golden rule is $1\mu$s. The solid lines are the numerical results, and the dashed lines are the approximations in Eq.~(\ref{eq:relax_fit}) and  Eq.~(\ref{eq:relax_ft_fit}).}\label{fig:relax}
\end{figure}

The simulation of both dephasing and relaxation present can be done easily in our program. The newly added spins remain non-interactive among themselves. Fig.~(\ref{fig:relax}) shows a simulation result with the relaxation time $T_1\gg T_\phi$\cite{gammat_phi}. In the real time evolution graph we can clearly see the three stages of the decay process, which starts with a fast Gaussian decay followed by slow decay, and then later the exponential decay finally takes over. Similar behavior has also been observed in the result produced by a qubit under the direct influence of 1/f noise\cite{Falci04}. In the graph of the Fourier transform of the same data, the original sharp peak in the Rabi frequency is now smeared. Since this problem cannot be solved analytically, we made the approximation of multiplying the oscillating part of Eq.~(\ref{eq:sz}) with an exponential factor $e^{-t/2T_1}$, so that it becomes
\begin{eqnarray}
\langle\sigma^z(t)\rangle & = & -1+\frac{\alpha^2}{\sqrt{2\pi}\delta\Omega}\label{eq:relax_fit}\\
& & \quad\cdot\int_{-\infty}^\infty dB'\frac{1-e^{-t/2T_1}\cos2\sqrt{\alpha^2+B'^2}t}{\alpha^2+B'^2}e^{-\frac{B'^2}{2\delta\Omega^2}}.\nonumber
\end{eqnarray}
The Fourier transform of this new function is
\begin{equation}
\tilde{\sigma}^z(\omega)\sim\int_{-\infty}^\infty\frac{1}{-i(\omega-2\sqrt{\alpha^2+B^2})+\frac{1}{2T_1}}\frac{e^{-\frac{B^2}{2\delta\Omega^2}}}{\alpha^2+B^2}dB,\label{eq:relax_ft_fit}
\end{equation}
and $\mathrm{Re}[\tilde{\sigma}^z(\omega)]$ has been used in approximating the numerical result in Fig.~(\ref{fig:relax}). Although the approximation deviates from the numerical data significantly in the frequency domain, which is particularly obvious in the tails of the peak, they are in reasonable agreement in the time domain. This approach thus allows us to easily extract the relevant physical quantities such as $T_\phi$ and $T_1$ from experimental data in a system of such complexity.

\section{Coupling to Spurious Resonators}
Recently experiment by Simmonds \emph{et. al}\cite{Simmonds04} reported the observation of qubit coupling to some spurious resonators. When the qubit is in resonance with one of these resonators, the Rabi oscillation pattern has a significant decrease in amplitude with irregular shape but not in decay time. In general, when a qubit couples with another two-level system while undergoing the coherently driven Rabi oscillation at its original frequency, it should display beats in the pattern\cite{Ku04,Falci04} instead of a decrease in the amplitude. As previously discussed, the reduction of amplitude is due to dephasing, thus we suspect the reason for the amplitude reduction is that the spurious resonator itself is also being dephased, which further decreases the amplitude and smears the beats.

The theoretical model we consider is as follows,
\begin{eqnarray}
\mathcal{H} & = & \mathcal{H}_\mathrm{qb}+\mathcal{H}_\mathrm{res}+\mathcal{H}_\mathrm{spins}+\mathcal{H}_\mathrm{int}+\mathcal{H}_\mathrm{Rabi};\nonumber\\
\mathcal{H}_\mathrm{qb} & = & \frac{\Omega}{2}\sigma^z,\nonumber\\
\mathcal{H}_\mathrm{res} & = & \frac{\Omega_\mathrm{res}}{2}\sigma_\mathrm{res}^z,\nonumber\\
\mathcal{H}_\mathrm{spins} & = & \sum_k\frac{\omega_k}{2}\tau_k^z,\nonumber\\
\mathcal{H}_\mathrm{int} & = & g(\sigma^+\sigma_\mathrm{res}^-+\sigma^-\sigma_\mathrm{res}^+)\nonumber\\
& & +\sum_kA_k\tau_k^z\sigma^z+\sum_kC_k\tau_k^z\sigma_\mathrm{res}^z,\nonumber\\
\mathrm{H}_\mathrm{Rabi} & = & \alpha(\sigma^+e^{-i\Omega't}+\sigma^-e^{i\Omega't}),
\end{eqnarray}
where $\Omega_\mathrm{res}$ denotes the frequency of the two-level resonator, which couples to the qubit with coupling constant $g$. The resonator is also dephased by the same group of spins that dephase the qubit. Here we drop the relaxation effect, because it is less significant. As in the single qubit case, this model is exactly solvable in the limit of an infinite number of spins. After applying the unitary transformation
\begin{equation}
U=\mathrm{exp}\left(i\frac{\Omega't}{2}\sigma^z+i\frac{\Omega't}{2}\sigma_\mathrm{res}^z+i\sum_k\frac{\omega_kt}{2}\tau_k^z\right),
\end{equation}
the Hamiltonian becomes
\begin{eqnarray}
\tilde{\mathcal{H}} & = & \frac{\Omega-\Omega'}{2}\sigma^z+\frac{\Omega_\mathrm{res}-\Omega'}{2}\sigma_\mathrm{res}^z+g(\sigma^+\sigma_\mathrm{res}^-+\sigma^-\sigma_\mathrm{res}^+)\nonumber\\
& & +\sum_kA_k\tau_k^z\sigma^z+\sum_kC_k\tau_k^z\sigma_\mathrm{res}^z+\alpha\sigma^x.
\end{eqnarray}
Because it is the renormalized frequencies of the qubit and the spurious resonator that are in resonance at the driving microwave frequency $\Omega'$, we can obtain an effective Hamiltonian by defining $B_\mathrm{res}\equiv\sum_kC_k\tau_k^z$ and $B_\mathrm{res}'\equiv B_\mathrm{res}-\langle B_\mathrm{res}\rangle$, to obtain
\begin{equation}
\tilde{\mathcal{H}}=g(\sigma^+\sigma_\mathrm{res}^-+\sigma^-\sigma_\mathrm{res}^+)+B'\sigma^z+B_\mathrm{res}'\sigma_\mathrm{res}^z+\alpha\sigma^x.
\end{equation}
This is a $4\times4$ matrix that can be easily diagonalized, though the result is a little messy. By the same token, physical quantities such as $\langle\sigma^z(t)\rangle$ can be calculated by integrating out all possible values of $B'$ and $B_\mathrm{res}'$.

The density matrix of the initial state is chosen so that both the qubit and the spurious resonator begin with the $|\downarrow\rangle$ state,
\begin{equation}
\rho_0=|\downarrow\rangle\langle\downarrow|_\mathrm{qb}\otimes|\downarrow\rangle\langle\downarrow|_\mathrm{res}\otimes e^{-\beta\sum_k\frac{\omega_k}{2}\tau_k^z}.
\end{equation}
The analytic result of the Fourier transform is approximately
\begin{widetext}
\begin{eqnarray}
\tilde{\sigma}^z(\omega) & \sim & \int dt\int_{-\infty}^\infty dB'\int_{-\infty}^\infty dB_\mathrm{res}'r(B',B_\mathrm{res}')e^{-\frac{B'^2}{2\delta\Omega^2}-\frac{B_\mathrm{res}'^2}{2\delta\Omega_\mathrm{res}^2}}e^{i[\omega\pm f(B',B_\mathrm{res}')]t},\\
f(B',B_\mathrm{res}') & = & 2\sqrt{\alpha^2+B'^2+B_\mathrm{res}'^2+g^2/2\pm\sqrt{g^4/4-2B'B_\mathrm{res}'g^2+4B'^2B_\mathrm{res}'^2+g^2\alpha^2+4B_\mathrm{res}'^2\alpha^2}},\label{eq:sz_ft_res}
\end{eqnarray}
\end{widetext}
where the definition of $\delta\Omega_\mathrm{res}$ is similar to Eq.~(\ref{eq:b_beta}), which is the energy fluctuation of the spurious resonator. The function $r(B',B_\mathrm{res}')$ is the other part of the integrand that does not depend on $\delta\Omega$, $\delta\Omega_\mathrm{res}$, or $t$. The variables $B'$ and $B_\mathrm{res}'$ are in fact not independent as suggested in the way the integral is written, which is only for the purpose of easier explanation.

In the case when $\delta\Omega_\mathrm{res}\ll \delta\Omega$, we can see that in the frequency exponent $f(B',B_\mathrm{res}')$ in Eq.~(\ref{eq:sz_ft_res}), the first three terms in the square root are similar to the frequency exponent in Eq.~(\ref{eq:FTsz}) besides an extra term $B_\mathrm{res}'^2$. This means that the Rabi frequency fluctuation is slightly increased by the fact that the resonator is being dephased. The rest of the terms in the square root are causing the beats. If we freeze $B_\mathrm{res}'=0$ for these terms, they would not fluctuate, and the oscillation would display a clear beat pattern. But if $B_\mathrm{res}'$ is allowed to fluctuate, the beat frequency will also fluctuate, which smears the beat pattern. This suggests that the disappearance of beats is due to the dephasing of the spurious resonator itself when $\delta\Omega_\mathrm{res}>g$. However, notice that the argument above applies only when the resonator fluctuates mildly, $\delta\Omega_\mathrm{res}<\delta\Omega$; otherwise the resonator would be brought out of resonance with the qubit.

\begin{figure}
\centering
\includegraphics[scale=0.6]{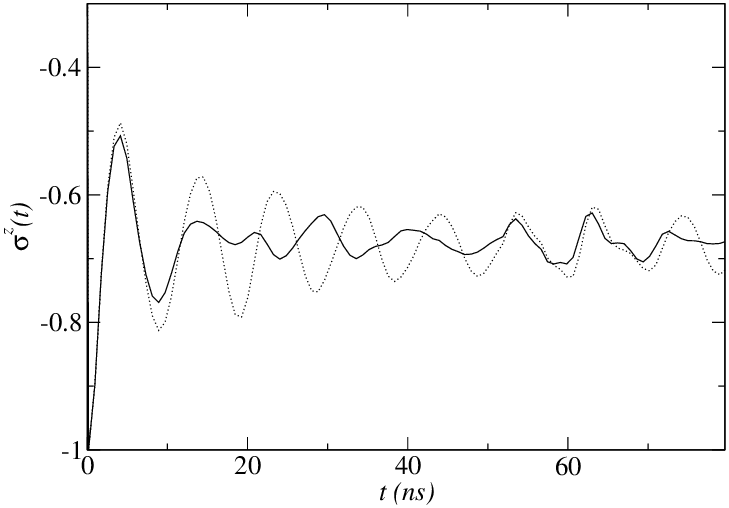}
\caption{Effect on Rabi oscillations caused by a dephased spurious resonator. The solid line represents the case where the qubit is coupled to a spurious resonator, and the dotted line is not. The parameters for both are $2\alpha/2\pi=100$MHz and $\delta\Omega/2\pi=145$MHz. Those for the coupled resonator are $g/2\pi=15$MHz and $\delta\Omega_\mathrm{res}/2\pi=30$MHz. Both qubit and spurious resonator couple to the same 14 spins. Notice that because $\delta\Omega>h$, the initial reduction of amplitude ends even before one period of Rabi oscillation.}\label{fig:spurious}
\end{figure}
Fig.~(\ref{fig:spurious}) shows the comparison of the numerical result of a qubit dephased by 14 spins with and without a spurious resonator, which is also dephased by the same group of spins. We can see that the Rabi oscillation becomes somewhat irregular with reduced amplitude while the long-time slow decay still persists without the sign of fading away. This explains the experimental observation\cite{Simmonds04} that decoherence time is not reduced by the coupling to the spurious resonator is because the time has already passed $T_\phi$ and the Rabi oscillations have entered the featureless slow decay regime. Notice that the irregularity starts to appear only after $t\sim2\pi/g$, which suggests the cause is due to the smearing of the beats.

\begin{figure}
\centering
\includegraphics[scale=0.6]{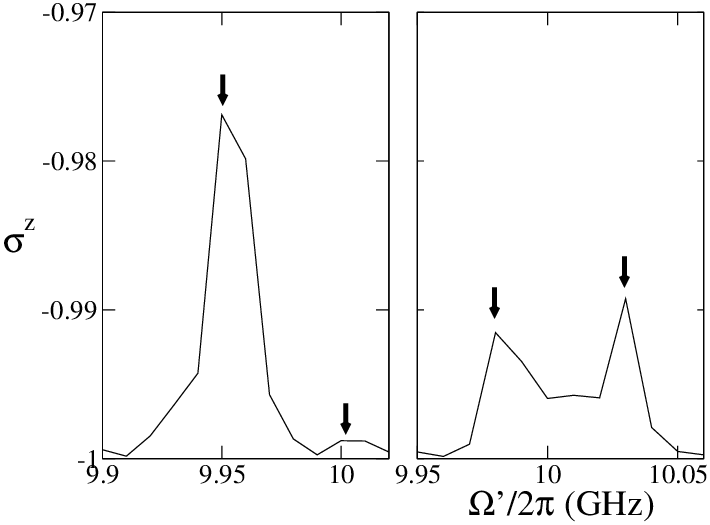}
\caption{Simulation of the spectral probing of resonance on the qubit-spurious resonator system. In both graphs, the vertical axes are the value of $\langle\sigma^z\rangle$ at the steady state, and the horizontal axes are the driving microwave frequencies. The spurious resonator here has an energy splitting at $10$GHz. The left graph shows when the qubit energy is detuned from the resonator at $9.95$GHz, the most visible peak is the qubit and the resonator peak is barely visible. When the qubit energy is tuned close to the resonator frequency, as shown in the right graph, level repulsion takes place. Notice that the spectral peaks here are clearly seperated even though $\delta\Omega$ and $\delta\Omega_\mathrm{res}$ are both greater than $g$.}\label{fig:gsplit}
\end{figure}
However, with the parameters used in Fig.~(\ref{fig:spurious}) where $\delta\Omega$ and $\delta\Omega_\mathrm{res}$ are both greater than $g$, one might suspect that since the fluctuations of the energy level splittings are strong enough to smear the beat pattern in the Rabi oscillations, they could also destroy the avoided crossings observed in the spectroscopic data of the qubit transition frequency\cite{Simmonds04}. Here this situation is numerically simulated and the result is shown in Fig.~(\ref{fig:gsplit}), where the parameters are mostly the same as those in Fig.~(\ref{fig:spurious}), except a much weaker driving microwave power of $2\alpha/2\pi=1$MHz to ensure that the probability of excitation to the up state is linear to the driving power and additional relaxation spins for faster convergence. We show that the spectral peaks in the simulation are still clearly seperated and that the dephasing by many two-level systems is fundamentally different from those by harmonic baths.

\section{Conclusion}
We have analytically and numerically solved the model of a qubit undergoing Rabi oscillation while being dephased by a non-interacting many-spin system. It is found that the oscillation pattern, in the short time limit, has a Gaussian decay, because the spins have no interaction among themselves. In the long time limit, it is fundamentally different from a qubit coupled to a heat bath\cite{Smirnov03}, for its longer than exponential decay.

The Fourier transform of $\langle\sigma^z(t)\rangle$ has a special shape and thus requires a definition of its spectral width. We define the dephasing time $T_\phi$ in Eq.~(\ref{eq:t_phi}) to quantify the simple relation between dephasing time, the Rabi frequency, and the fluctuation of qubit energy splitting. We found that the dephasing by a many-spin system governs the initial amplitude reduction of the Rabi oscillation, which ends around $t\sim T_\phi$. The slow decay later is a general behavior of a qubit coupled to a many-spin system.

When the relaxation mechanism is added to the system, numerical simulation shows that one more stage of exponential decay appears after the slow decay in the pure dephasing case. Since this model cannot be solved exactly, we established an approximation method to allow for easy extraction of physical quantities of decoherence times from experimental data.

We have also numerically reproduced the serious reduction of Rabi oscillation amplitude caused by a spurious resonator by considering the situation that the spurious resonator is also dephased by the same group of fictitious spins. The reduced amplitude of oscillation and the persistence of slow decay are both demonstrated.

\section{Acknowledgment}
The authors would like to thank Steven M. Girvin for discussions. This work was supported by the National Security Agency (NSA) and Advanced Research and Development Activity (ARDA) under Army Research Office (ARO) contract DAAD 19-03-1-0132. This work was partially carried out at the Ames Laboratory, which is operated for the U.~S. Department of Energy by Iowa State University under Contract No.~W-7405-82 and was supported by the Director of the Office of Science, Office of Basic Energy Research of the U.~S. Department of Energy.

\bibliography{t-dep}

\begin{thebibliography}{15}
\bibitem[{gam()}]{gammat_phi}
\bibinfo{note}{This condition is chosen, because for solid state qubits usually
  $T_2$ is much shorter then $T_1$.}
\end{thebibliography}

\end{document}